\documentclass{PoS}

\def\cmss{\mathrm{cm}^{-2}\;\mathrm{s}^{-1}}
\def\lumiLOW{10^{23}\;\cmss}
\def\lumiLow{10^{25}\;\cmss}
\def\xlumiLOW{\times\lumiLOW}
\def\xlumiLow{\times\lumiLow}

\def\bstar{\beta^{\star}}
\def\bmax{\beta_{\mathrm max}}
\def\snn{\sqrt{s_{NN}}}

\def\SCRUNCH{\vspace*{-3mm}}

\def\Unscrunch{\vspace*{2mm}}

\newcommand\T{\rule{0pt}{2.6ex}}
\newcommand\B{\rule[-1.2ex]{0pt}{0pt}}

\title{RHIC Low-Energy Challenges and Plans}
\ShortTitle{RHIC Low-Energy Challenges and Plans}
\author{\speaker{T.~Satogata}\thanks{Work supported by the US
Deparment of Energy under Contract No.~DE-AC02-98CH1-886.}, L.~Ahrens,
M.~Bai, J.M.~Brennan, D.~Bruno, J.~Butler, A.~Drees, A.~Fedotov,
W.~Fischer, M.~Harvey, T.~Hayes, W.~Jappe, R.C.~Lee, W.W.~MacKay,
N.~Malitsky, G.~Marr, R.~Michnoff, B.~Oerter, E.~Pozdeyev, T.~Roser,
F.~Severino, K.~Smith, S.~Tepikian, and N.~Tsoupas\\
        Brookhaven National Laboratory\\
        E-mail: \email{satogata@bnl.gov},
                \email{ahrens@bnl.gov},
                \email{mbai@bnl.gov},
                \email{brennan@bnl.gov},
		\email{bruno@bnl.gov},
		\email{jbutler@bnl.gov},
		\email{drees@bnl.gov},
		\email{fedotov@bnl.gov},
		\email{wfischer@bnl.gov},
		\email{pharvey@bnl.gov},
		\email{hayes@bnl.gov},
		\email{jappe@bnl.gov},
		\email{rclee@bnl.gov},
		\email{waldo@bnl.gov},
		\email{malitsky@bnl.gov},
		\email{gmarr@bnl.gov},
		\email{michnoff@bnl.gov},
		\email{oerter@bnl.gov},
		\email{pozdeyev@bnl.gov},
		\email{roser@bnl.gov},
		\email{severino@bnl.gov},
		\email{ksmith@bnl.gov},
		\email{tepikian@bnl.gov},
		\email{tsoupas@bnl.gov}
}
\abstract{There is significant interest in RHIC heavy ion collisions
at $\snn=$5--50 GeV, motivated by a search for the QCD phase
transition critical point. The lowest energies for this search are
well below the nominal RHIC gold injection collision energy of
$\snn=19.6$ GeV. There are several operations challenges at
RHIC in this regime, including longitudinal acceptance, magnet field
quality, lattice control, and luminosity monitoring. We report on the
status of work to address these challenges, including results from
beam tests of low energy RHIC operations with protons and gold, and
potential improvements from different beam cooling scenarios.}

\FullConference{Critical Point and Onset of Deconfinement - 4th International Workshop\\
		 July 9 - 13, 2007\\
		 Darmstadt, Germany}

\begin{document}
\section{BACKGROUND AND MOTIVATION}
There is significant theoretical and experimental evidence that points
to the existence of a QCD phase transition critical point on the QCD
phase diagram. If this critical point exists, it should appear on the
quark-gluon phase transition boundary in the range of baryo-chemical
potential of 100--500 MeV \cite{REF:ORIG}. This corresponds to heavy
ion collisions at RHIC with $\snn=$5--50 GeV. Experimental
identification of this critical point would be a major step towards
the understanding of QCD at high temperatures and densities.

Experimental exploration of heavy ion collisions in this energy regime
is feasible using the STAR and PHENIX detectors at RHIC. This data
would complement existing fixed-target data from the AGS (2.5--5 GeV)
and SPS (5--20 GeV). The required integrated luminosities for this
search are low and challenging -- approximately $5\times10^6$ min-bias
events are needed at each of 6--7 energies to improve on existing NA49
statistics by a factor of 2--4 \cite{REF:RIKEN,REF:STEPHANS}.

\begin{figure}[b]\centering
\includegraphics*[width=90mm]{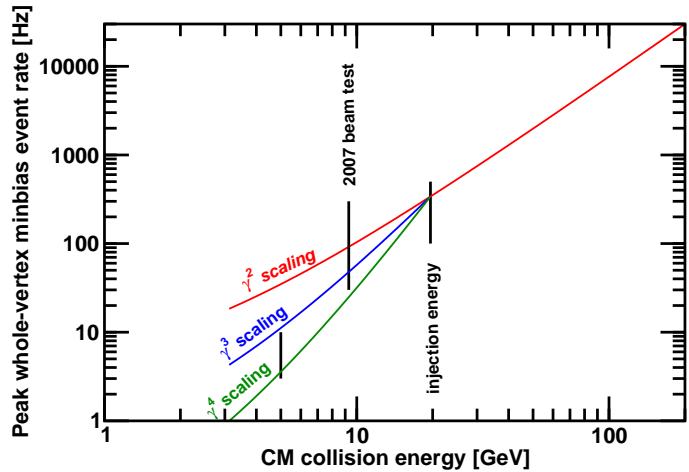}
\caption{Scaling of RHIC minbias event rate (or luminosity) into the
  low energy regime. Energies of interest are $\snn=$5--50 GeV,
  or beam kinetic energies of 1.6--24.1 GeV/nucleon. Vertical lines are
  measured ($\snn=19.2$~GeV at injection; $\snn=9.18$~GeV in the 2007
  beam test) and projected ($\snn=5$~GeV) minimum to maximum event rates.}
\label{FIG:SCALING}
\end{figure}

Fig.~\ref{FIG:SCALING} shows several scalings for RHIC Au-Au
luminosity in the low-energy regime of interest. Above the nominal
injection energy of 9.8 GeV/nucleon, the beam size and aperture both scale
with $\gamma$, so the event rate (or luminosity) scales as
$\gamma^2$. Here RHIC runs as a ramping collider. Previous collider
runs at several intermediate energies are consistent with prediction,
with present peak luminosity $L_{\mathrm peak}=4.0\xlumiLow$ at
injection energy and $L_{\mathrm peak}=3\times10^{27}\;\cmss$ at
$\snn=200$~GeV.

At and below nominal injection energy, RHIC runs as a colliding
storage ring, and beam size, field quality, longitudinal acceptance,
and intrabeam scattering (IBS) growth will conspire to make luminosity scaling
worse. Fig.~\ref{FIG:SCALING} shows $\gamma^3$ and $\gamma^4$ scalings
as examples. The 2007 Au beam test, detailed in section 4 of this
paper, demonstrated better than $\gamma^2$ scaling at
$\snn=9.18$~GeV. However, performance will certainly degrade at lower
energies; an estimate of event rates for $\snn=5$~GeV corresponding to $L_{\mathrm
peak}\approx5\times10^{22}\;\cmss$ is also shown.

\section{PARAMETERS}
Table~\ref{TAB:PARAMS} compares some RHIC parameters that are relevant for
low-energy operations, including two test runs that occurred in 2006
with protons and 2007 with gold. Summaries of these test runs are presented
in the sections 3 and 4 of this paper.

\begin{table*}[t]\centering
\caption{Parameters for nominal RHIC Au injection, 2006 and 2007 low-energy
         test runs with protons and gold, and the lowest energy of interest
         for the QCD critical point search. Injection and 2007 test run
         peak luminosities are measured; the $\snn=$5~GeV peak luminosity
         is estimated. Beam sizes are
         calculated assuming $\epsilon_{\mathrm N}$(Au)=40$\pi\;\mu$m,
         $\epsilon_{\mathrm N}$(p)=10$\pi\;\mu$m, $\bstar$=10m, and $\bmax$=170m.
         $L_{\mathrm peak}$ assumes $\bstar$=10m.}
\Unscrunch
\begin{tabular}{ccccccccccc} \hline
\T Species   & $\snn$ & $KE_{\mathrm beam}$ & $\gamma$ & B$\rho$ & $f_{\mathrm rev}$  & h & $\sigma^{*}_{\mathrm 95\%}$ & $\sigma_{\mathrm max, 95\%}$ & $L_{\mathrm peak}$\\
\B           & [GeV]    & [GeV/nucleon]         &          & [T-m]   & [kHz]          &     & [mm]                    & [mm] & $[\cmss]$ \\ \hline
\T Au (inj)  &  23.47     & 10.80           & 12.6     & 97.3    & 77.95          & 360 & 2.3 & 9.5 & 400$\times10^{23}$ \\
p (2006)     &  22.5      & 10.31           & 11.99    & 37.4    & 77.92          & 360 & 1.2 & 4.9 &  -- \\
Au (2007)    &  9.18      & 3.66            & 4.93     & 37.4    & 76.57          & 366 & 3.7 & 15.3 & 4$\times10^{23}$ \\
\B Au (low)  &  5.0       & 1.57            & 2.68     & 19.3    & 72.57          & 387 & 5.2 & 21.3 & 5$\times10^{22}$ \\
\hline\SCRUNCH
\label{TAB:PARAMS}
\end{tabular}
\end{table*}
For linear field response, power supply current scales with magnet
rigidity B$\rho$. At the lowest requested collision energy, rigidity
and power supply currents are only 20\% of their values at nominal
injection energy. Main power supply regulation has been tested in RHIC
at these currents, and shows no problems. Other field quality issues were
experimentally investigated during the 2006 and 2007 test runs and also
show no serious concerns at $B\rho=$37.4~T-m.

\begin{table*}[b]\centering
\caption{RF constraints for RHIC low-energy operations. Even harmonic numbers
  may be required for RHIC instrumentation and beam sync clock issues; RF software
  currently requires harmonic numbers divisible by 9 to generate simultaneous
  collisions at both experiments. Eliminating both these constraints would allow
  operation at any harmonic number divisible by 3.}
\Unscrunch
\begin{tabular}{cccccc} \hline
\T Harmonic number & $\snn$ range & h(mod2)=0 & h(mod3)=0 & h(mod9)=0 & h(mod18)=0 \\ 
\B      $h$        & [GeV]        &           &           &           &            \\ \hline
   360             & 16.7--107    &  *        & *         & *         & * \\
   363             & 11.4--15.0   &           & *         &           &  \\
   366             & 9.0--10.5    &  *        & *         &           &  \\
   369             & 7.7--8.6     &           & *         & *         &  \\
   372             & 6.9--7.4     &  *        & *         &           &  \\
   375             & 6.3--6.7     &           & *         &           &  \\
   378             & 5.8--6.1     &  *        & *         & *         & * \\
   381             & 5.45--5.7    &           & *         &           &  \\
   384             & 5.15--5.38   &  *        & *         &           &  \\
   387             & 4.91--5.1    &           & *         & *         &  \\
\hline\SCRUNCH
\label{TAB:HARMONICNUMBER}
\end{tabular}
\end{table*}

At low energies, Au beam becomes less relativistic, and the ion beam
RF frequency is lowered out of the RHIC RF tuning range of
28.0--28.17~MHz for the standard RHIC harmonic number $h=360$. The
harmonic number must therefore be raised for collision energies less
than $\snn=16.7$~GeV. RHIC is three-fold symmetric, so only
harmonic numbers divisible by 3 can produce simultaneous collisions at
both STAR and PHENIX experiments. RF software related to injection patterns
currently requires harmonic numbers divisible by 9 to produce collisions
at both experiments, and RHIC beam synchronous clock systems \cite{REF:BEAMSYNC}
currently require harmonic numbers divisible by 2 to operate properly. Work
is underway to eliminate both these constraints and permit operation at any
harmonic number divisible by 3, as shown in Table~\ref{TAB:HARMONICNUMBER}.

Longitudinal and transverse acceptance vs emittance is another
challenge at low energies. RHIC Au beam typically has a longitudinal
emittance of 0.2~eV-s/u at injection. This beam barely fits into the RHIC
bucket RF with 400~kV at $\snn=9$~GeV. At $\snn=5$~GeV
longitudinal acceptance is only 0.12~eV-s/u, and 30--50\% of the beam
immediately debunches even with perfect longitudinal
injection. Transverse acceptance issues in the transfer line provide
similar limitations, leading to expectations of only 20--40\%
injection efficiency at the lowest energy.

\section{2006 PROTON TEST RUN}
The first 24-hour test of RHIC at low energy occurred June 5--6 2006,
during the 2006 RHIC polarized proton run. As Au beam was not
available, the objective of this run was to evaluate setup time, power
supply behavior, linear field quality, beam stability, and
optics. The RHIC rigidity was $B\rho$=37.4~T-m, corresponding to beam kinetic
energy of 10.31~GeV, less than half of the nominal proton injection
kinetic energy of 22.87~GeV. This rigidity was chosen because it
corresponds to a potentially interesting feature in the QCD phase
diagram \cite{REF:STEPHANS}; it is also about halfway between nominal
injection and the lowest energy of interest. Protons are still highly
relativistic at this energy so the RHIC harmonic number was
unchanged.

After initial setup, first circulating beam was achieved in
approximately 3 hours in both rings; another 3 hours were needed for
RF setup and capture.  Injection efficiencies were 70--80\% with beam
lifetimes of 5--10 hours at the normal polarized proton working
point. Vernier scan attempts did not provide good data, due to lack of a
clean luminosity signal and high backgrounds from unstable beam.

Optics measurements were performed using difference orbits and orbit
response matrices. These measurements indicated 10-15\% beta waves in
both planes compared to the design injection model, consistent with
optics quality at nominal injection field in RHIC. This combined with
excellent beam lifetimes indicated that field quality at these
energies was not problematic. Dipole ripple was also measured with beam, and
showed no deviation from nominal injection spectra; power supply
ripple and regulation at this energy is therefore also not an issue.

Low-field extrapolation of RHIC magnet measurements shows that the
RHIC main dipole sextupole component nearly doubles from -9 units to
-16 units from nominal injection to this energy. This created
additional chromaticity that was not fully compensated with the
existing configuration of RHIC sextupoles. Vertical chromaticities
could only be set to about +1 to +2 below transition energy instead of
the desired value of -1; instabilities were damped with strong
octupoles, though beams continued to be metastable. For physics runs
at this and lower energy, defocusing sextupole power supplies will be
reversed to allow proper chromaticity control without compromising
dynamic aperture.

\begin{figure}[b]\centering
\includegraphics[width=120mm]{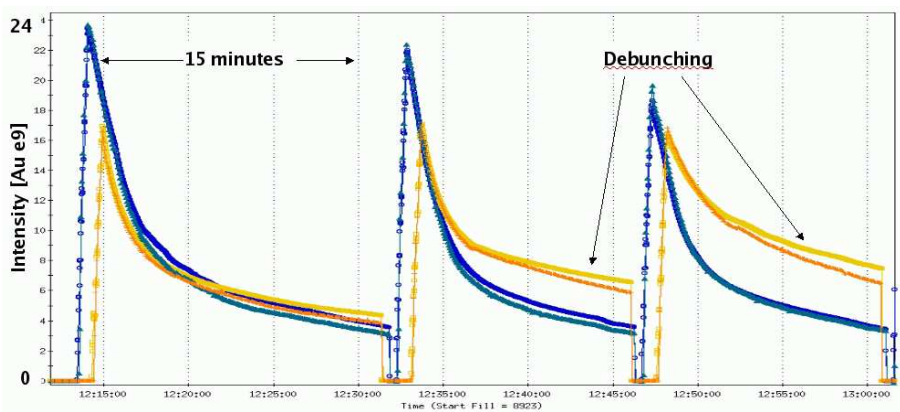}
\caption{Au beam lifetime at $\snn=9.18$ GeV (blue and yellow rings)
  during the 2007 gold test run, with
  a slow exponential decay component of 20 minutes and a fast
  exponential decay component of 2 minutes. Bunched and total beam
  currents are displayed; debunching is clearly visible, so momentum
  aperture is larger than the RF bucket size.}
\label{FIG:LIFETIME}
\end{figure}

\section{2007 GOLD TEST RUN}
The second 24-hour test of RHIC at low energy occurred June 6--7 2007,
during the 2007 RHIC Au-Au run. The RHIC rigidity was
$B\rho$=37.4~T-m, the same as the proton test run, to leverage the
2006 test run setup. This corresponded to an Au beam kinetic energy of
3.66 GeV/nucleon. The objectives of this run were to use a new RHIC harmonic
number, evaluate Au acceptances, and
measure Au-Au luminosity to place a measured low-energy point on
Fig.~\ref{FIG:SCALING}.

$h=366$ setup was straightforward for RHIC RF and AGS to RHIC synchro.
The RHIC beam synchronous event system also relies on an RF clock to
generate experiment trigger clocks and other beam-synchronous RHIC
instrumentation timing \cite{REF:BEAMSYNC}. Event generator hardware
that generates its own $h=360$ revolution fiducial event was bypassed
and the beam synchronous links were reconfigured, but at the cost of
priority of the fiducial event on the link. PHENIX could not lock to
this clock, but STAR could, with trigger resets every few minutes.

Fig.~\ref{FIG:LIFETIME} shows Au beam lifetime for three consecutive
stores during this low-energy run. Injection efficiencies were 70--80\%.
Decomposition of the beam lifetime shows two main exponential components:
a slow component of 20 minutes, and a fast component of 2 minutes. The
slow component is consistent with an IBS growth time prediction at this
energy. Measured transverse emittances were $\epsilon_{N,x,y}=15-25\;\pi\;\mu$m.
Longitudinal injection efficiency was 100\%; estimated longitudinal
emittance was 0.14 eV-s/nucleon, significantly smaller than the expected
0.2 eV-s/nucleon, perhaps because AGS transition crossing was unnecessary.

\begin{figure}[t]\centering
\includegraphics[width=90mm]{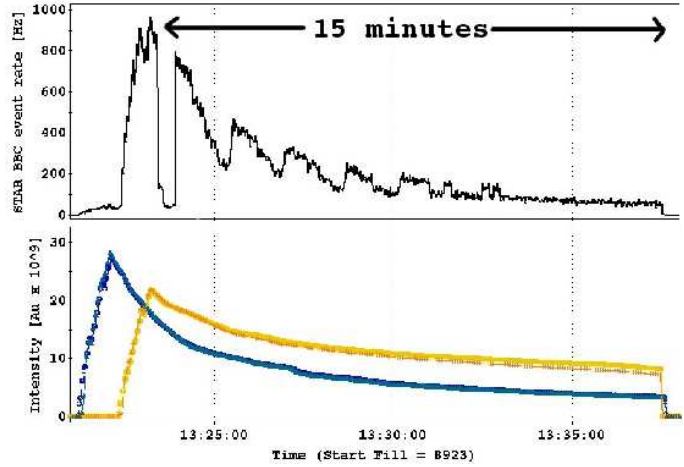}
\caption{A 15-minute Au-Au store at $\snn=9.18$ GeV, showing blue
  and yellow beam intensities and STAR BBC counter collision rates.
  Beams were uncogged and recogged at the start of the store; vernier
  scans in both planes were performed during the store.}
\label{FIG:VERNIERSCANBEAM}
\end{figure}

Four vernier scans were acquired at the STAR experiment. Unfortunately
none could be acquired with new PHENIX detectors due to trigger clock
problems.  Fig.~\ref{FIG:VERNIERSCANBEAM} shows a STAR vernier scan
over $\pm9$ mm in 15 minutes in the horizontal planes. Beams were longitudinally
cogged out of collision early in the store, demonstrating only 5\%
backgrounds, though the gaussian baseline in the figure appears higher
than this. The beam average $\sigma$ of 5.98 mm is consistent with
a minimum $\sigma$ of 4 mm and an average normalized
horizontal emittance of 25 $\pi\;\mu$m, giving a measured peak luminosity
of $4\xlumiLOW$ as listed in Table~\ref{TAB:PARAMS}.

\begin{figure}[t]\centering
\includegraphics[width=55mm]{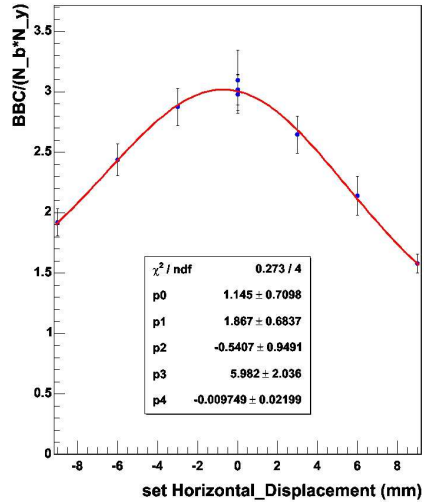}
\caption{A STAR BBC vernier scan at $\snn=9.18$ GeV, scanning beam
  horizontally over $\pm$6 mm. There was little background
  contamination, in contrast to the 2006 proton test run.}
\label{FIG:VERNIERSCAN}
\end{figure}

\section{PROJECTIONS AND COOLING PROSPECTS}
Table~\ref{TAB:PROJECTIONS} shows a strawman proposal for a basic
low-energy run exploring the energies from $\snn=4.6$--28~GeV, or
baryo-chemical potential $\mu_{\mathrm B}=150$--570~MeV as presented
by T.~Nayak at the 2006 RIKEN workshop \cite{REF:RIKEN}. The
parenthesized values are modified projections based on 2007 test run
results, which showed that 300~Hz average event rates were achievable
at $\snn=9$~GeV.  Projections below $\snn=8.8$~GeV have progressively
larger uncertainties due to unknown dynamic aperture and luminosity
lifetime. Even so, this indicates that a basic QCD critical point scan
might be completed at RHIC in a six-week run.

Thorough measurements of $K/\pi$ ratio fluctuations, jet quenching, and
other possible critical point signatures require larger data sets, on
the order of 10-50M events per energy point. Run time in Table~\ref{TAB:PROJECTIONS}
is dominated by run time of the lowest energy; a ten- to twelve-week RHIC run
would provide high statistics at all energies of interest. An optimal run
plan starts at the highest energies, providing scheduling flexibility if
promising signatures are measured at any intermediate energy.

Since the run time is dominated by the run time of the lowest energy,
run planning is dominated by uncertainties in the machine performance
at the lowest planned energy of $\snn=5$ GeV. A test of gold
collisions near this energy has been proposed for the 2008 RHIC run to
determine luminosity and luminosity lifetime, and to evaluate
requirements for potential AGS electron cooling. Injection efficiency
of 20--50\% and IBS lifetimes of a few minutes are expected, so
vernier scans and luminosity measurement will be challenging. Beam
synchronous clock issues for harmonic numbers other than 360 should be
resolved during the 2007 shutdown and tested with experiment triggers.

\begin{table}[t]\centering
\caption{A comparison of the STAR strawman low-statistic basic run proposal and extrapolations
  from the 2007 RHIC low energy test run (in parentheses). Beam days are expressed
  as run time plus setup time, assuming 50\% combined detector/facility uptime.
  Whole-vertex average BBC rates are listed , and do not include detector vertex acceptance.}
\Unscrunch
\begin{tabular}{cccccc} \hline
\T $\snn$  & $\mu_{\mathrm B}$ & <Minbias BBC Rate> & Days/     & \# events    & \# beam days \\
\B [GeV]   &  [MeV]            &   [Hz]             &  Mevent   & $\times10^6$ &              \\ \hline
   4.6     &   570             &  3 ($\sim$5)       & 9 (4.6)   & 5            & 45 (23+2)    \\
   6.3     &   470             &  7 ($\sim$50)      & 4 (0.5)   & 5            & 20 (3+1)     \\
   7.6     &   410             &  13 ($\sim$150)    & 2 (0.2)   & 5            & 10 (1+1)     \\
   8.8     &   380             &  20 (300)          & 1.5 (<1)  & 5 (>5)       & 7.5 (1+1)    \\
   12      &   300             &  54 ($\sim$1000)   & 0.5 (<1)  & 5 (>50)      & 2.5 (1+1)    \\
   18      &   220             &  >100 (>1000)      & 0.25 (<1) & 5 (>50)      & 1.5 (1+1)    \\
   28      &   150             &  >100 (>1000)      & 0.25 (<1) & 5 (>50)      & 1.5 (1+2)    \\
\hline\SCRUNCH
\label{TAB:PROJECTIONS}
\end{tabular}\end{table}

RHIC low energy integrated luminosity is primarily constrained by IBS
and field quality. Simulations indicate that initial gold beam IBS
growth rates at $\snn=5$~GeV are 250 and 100~s for transverse and
longitudinal emittances, respectively \cite{REF:COOL07}. In these
conditions optimal RHIC store lengths are 3--5 minutes. Optimal
normalized average luminosity $\langle L\rangle/L_{\mathrm peak}$ strongly depends
on the beam lifetime, ranging from 0.1--0.5 over beam lifetimes
of 30--180~s. 

Low-energy electron cooling in RHIC would counteract IBS beam loss,
improve beam capture, and permit long low-energy stores. Electron
cooling upgrades planned for the RHIC high-energy physics program
cannot be used here, since electron and ion velocities must match. One
possibility is ERL-based cooling, using a prototype half-cell
superconducting RF gun delivering bunched 1~nC electrons with kinetic
energies $E_{\mathrm k}=0.9-2.8$~MeV over a cooling length of 20~m.
Another option is to use a DC electron beam, such as that from the
Recycler cooler at Fermilab \cite{REF:PELLETRON}. Preliminary
simulations indicate that both systems would provide similar
performance, improving peak luminosity by 15--30 and integrated
luminosity by up to a factor of 100 \cite{REF:COOL07}.

For energies below $\snn=9$~GeV in the 2007 test run, the AGS
longitudinal emittance is likely too large to fit into the RHIC RF
acceptance. At the lowest energy, this can result in immediate
debunching of 50--80\% of injected beam. One way to counteract this is
with AGS injection energy electron cooling. Simulations indicate
that such a cooler could reduce gold beam longitudinal emittance by a
factor of 5, improving peak and integrated luminosity by a factor of
5--15. Integrated luminosity would not dramatically improve since
store lengths are still dominated by IBS lifetime in RHIC. This cooler
requires a cooling section length of ~1.5m, solenoidal field of ~0.1
T, electron energy of ~50 keV, and electron current of ~0.5A. These
parameters are easily achievable with existing technology and
expertise, but IBS and space charge limitations require careful study
\cite{REF:COOL07}. Longitudinal dampers are also being considered to
reduce longitudinal emittance growth at AGS injection.

\section{CONCLUSIONS}

RHIC heavy ion collisions at $\snn=$5--50 GeV are motivated by a
search for the QCD phase transition critical point. Two test runs,
with protons in 2006 and gold in 2007, have demonstrated program
feasibility at $\snn=9.18$ GeV. Gold beam parameters are better
than expected. RHIC harmonic number changes present minor problems;
most, including experiment clocking problems, will likely be fixed
during the 2007 summer shutdown. Projections estimate that a physics
program with 10--50M minbias events at each of seven energies, ranging
from $\snn=4.6$--28~GeV and $\mu_{\mathrm B}=150$--570~MeV, is
feasible in a 10--12 week RHIC run with no further improvements.  This
estimate is dominated by uncertainties in the lowest energy
performance. Optimal lowest energy store lengths are 3--5 minutes with
predicted peak luminosity of $5\times10^{22}\;\cmss$. RHIC and
AGS electron cooling are being studied as possible upgrade paths to
improve RHIC low-energy performance.

\section{ACKNOWLEGEMENTS}
The authors thank W.~Christie and M.~Leitch for experiment liason support,
and the RHIC/AGS operations staff for their continued support.

\end{document}